\documentclass[twocolumn,preprintnumbers,nofootinbib]{revtex4}
\usepackage{graphicx}
\usepackage{dcolumn}
\usepackage{bm}
\usepackage{amsmath,amssymb,amsfonts}
\usepackage{latexsym}
\usepackage{bbm}
\usepackage{url} 
\usepackage{color}
\begin{document}
\preprint{ACFI-T18-03}

\title{Electric Dipole Moments from CP-Violating Scalar Leptoquark Interactions}

\author{Kaori Fuyuto$^1$}
\email{kfuyuto@umass.edu}
\author{Michael Ramsey-Musolf$^{1,2}$}
\email{mjrm@physics.umass.edu}
\author{Tianyang Shen$^{1}$}
\email{shenty1991@gmail.com}
\affiliation{$^1$Amherst Center for Fundamental Interactions, Department of Physics,
University of Massachusetts Amherst, MA 01003, USA}
\affiliation{$^2$Kellogg Radiation Laboratory, California Institute of Technology, Pasadena, CA 91125 USA}
\bigskip

\date{\today}

\begin{abstract}
We analyze the implications of  CP-violating  scalar leptoquark (LQ) interactions  for experimental probes of parity- and time-reversal violating properties of polar molecules.
These systems are predominantly sensitive to the electric dipole moment (EDM) of the electron and nuclear-spin-independent (NSID) electron-nucleon interaction. The LQ model can generate both a  tree-level NSID interaction as well as the electron EDM at one-loop order.  Including both interactions, we find that the NSID interaction can dominate the molecular response. For moderate values of couplings, the current experimental results give roughly two orders of magnitude stronger limits on the electron EDM than one would otherwise infer from a sole-source analysis.
\end{abstract}

\maketitle

\section{Introduction}\label{sec:Intro} 

Explaining the origin of the cosmic matter-antimatter asymmetry requires violation of CP-invariance beyond that contained in the Standard Model (SM) of particle physics. Experimentally, among the most powerful probes of possible new CP-violating (CPV) interactions are searches for the permanent electric dipole moments (EDMs) of elementary particles and composite systems (for recent reviews, see Refs.~\cite{Pospelov:2005pr, Engel:2013lsa, Yamanaka:2017mef, Chupp:2017rkp}). 
The interaction of an electric field with the EDM of a quantum system violates both parity (P) and time-reversal (T) invariance,  and, assuming CPT invariance, implies the presence of CPV. 
Thus far, EDM searches have resulted in null results, yielding stringent upper bounds on the EDMs of the electron \cite{Baron:2013eja, Baron:2017, Cairncross:2017fip}, neutron \cite{Afach:2015sja} and mercury atom \cite{Graner:2016ses}. The most recent electron EDM bounds have been obtained from experiments using polar molecules:  $|d_e|<9.4\times 10^{-29}~e~{\rm cm}~(90\%~{\rm C.L.})$ \cite{Baron:2013eja, Baron:2017} ($^{232}$Th$^{16}$O)
and $|d_e|<1.3\times 10^{-28}~e~{\rm cm}~(90\%~{\rm C.L.})$ \cite{Cairncross:2017fip} ($^{180}$Hf$^{19}$F$^+$). 

Since electron EDM searches have yet to be performed using unbound electrons, experiments have thus far relied on paramagnetic systems. Polar molecules are particularly advantageous, as the unpaired electron experiences a significantly larger internal molecular electric field as compared to the applied external field. On the other hand, the signal associated with a non-vanishing $d_e$ can also be induced by a CPV interaction between the unpaired electron and the quarks in the nucleus, resulting in a nuclear-spin-independent (NSID) electron-nucleon interaction. The aforementioned $d_e$ bounds assume this NSID contribution vanishes. From a theoretical perspective, however, this \lq\lq sole source" assumption is not in general justified. The four-fermion CPV semileptonic interactions inducing the NSID interaction arise in well-motivated scenarios for physics beyond the Standard Model (BSM) \cite{Barr:1992cm, He:1992dc, Herczeg:2003ag}. Moreover, as pointed out in Ref.~\cite{Chupp:2014gka}, for a given level of experimental sensitivity, the resulting constraints on the BSM mass scale $\Lambda$ associated with the NSID contribution may be up to three orders of magnitude higher than those associated with the electron EDM. This difference results from three features: 
\begin{itemize}
\item The CPV SU(2$)_L\times$U(1$)_Y$-invariant $eq$ operator can be generated without any Yukawa interaction, whereas the electron EDM, being a dipole operator, requires a Yukawa insertion to ensure gauge-invariance, typically leading to a factor of $~ 10^{-6}$ suppression.
\item The CPV $eq$ operator can arise from the tree-level exchange of a BSM meditator, whereas the EDM first appears at one-loop order.
\item The NSID interaction samples the coherent sum of contributions from all nucleons, leading to an additional enhancement by a factor of the nuclear mass number.
\end{itemize}
These relatively model-independent considerations imply that the present polar molecule EDM bounds probe $\Lambda$ of order 1000 TeV (1 TeV) associated with the NSID ($d_e$) contributions. Thus, one expects the NSID effect will dominate the constraints on any CPV BSM scenario that generates both the $eq$ interaction and $d_e$. 

In what follows, we consider a concrete realization of this expectation in the context of leptoquark (LQ) scenarios
\cite{Buchmuller:1986zs, Davies:1990sc, Davidson:1993qk, Dorsner:2016wpm}. Leptoquarks have long been studied in particle physics and have recently received additional attention as possible explanations of $B$-physics anomalies \cite{Fajfer:2012jt,Dorsner:2013tla,Sakaki:2013bfa,Bhattacharya:2016mcc,Crivellin:2017zlb}. The LQ-quark-lepton coupling can be large and complex. The corresponding implications  for CP-violating processes involved in $K$-meson decays are discussed in \cite{Bobeth:2017ecx}. LQ models that allow couplings to both left- and right-handed quarks can accommodate the chiral flip needed to generate dipole dipole operators  (in the presence of the associated Yukawa interaction), a feature that has been analyzed for $d_e$  in Ref.~ \cite{Arnold:2013cva}.  Depending on the flavor structure of the LQ model, the CPV $eq$ interactions may also be induced at tree level \cite{Barr:1992cm, He:1992dc, Herczeg:2003ag}. In this case,  the NSID interaction becomes a dominant source of the P- and T-odd effect in paramagnetic systems. We illustrate these possibilities by considering two cases: (a) a flavor diagonal LQ model, wherein both $d_e$ and the tree-level CPV electron-up quark interaction arise; (b) a flavor non-diagonal scenario involving first and third generation fermions. 

The paper is organized as follows. First, we briefly review the polar molecule system sensitivity to low-energy CP-violating interactions in Sec. \ref{sec:molecule}. In Sec \ref{sec:model}, the LQ-induced $d_e$ and the CPV $eq$  interactions are derived. We analyze the corresponding implications for the interpretation of EDM experiments in Sec. \ref{sec:results}. Here, we also comment on the sensitivities of the neutron and proton EDMs to the LQ-induced quark EDM and chromo-EDM operators. We conclude in Section \ref{sec:conclusion}.

\section{CP violation in polar molecule system}
\label{sec:molecule}
The response of paramagnetic polar molecules to an applied external electric field is dominated by the electron EDM and the NSID electron-nucleon interactions. The electron EDM interaction is given by
\begin{align}
{\cal L}^{\rm EDM}=-\frac{i}{2}d_e\bar{e}\sigma^{\mu\nu}\gamma_5eF_{\mu\nu}, \label{de}
\end{align}
where $F_{\mu\nu}$ is the electromagnetic field strength. The NSID electron-nucleon interaction is described by
\begin{align}
{\cal L}^{\rm NSID}_{eN}=-\frac{G_F}{\sqrt{2}}\bar{e}i\gamma_5e~\bar{\psi}_N\left(C^{(0)}_S+C^{(1)}_S\tau^3 \right)\psi_N, \label{NSID}
\end{align}
with the Fermi constant $G_F$, a nucleon spinor $\psi_N$, and the Pauli matrix $\tau^3$.
Defining the following combination
\begin{align}
C_S\equiv C_S^{(0)}+\frac{Z-N}{Z+N}C^{(1)}_S
\end{align}
with the proton $Z$ and neutron $N$ numbers, the frequency associated with the  polar molecule response is \cite{Baron:2017}
\begin{align}
\omega=-E_{\rm eff}d_e+W_SC_S.
\end{align}
Note that $C_S$ is dominated by $C_S^{(0)}$ since $Z\sim N\sim O(100)$ for the systems of experimental interest. The field $E_{\rm eff}$ is called an effective electric field, and  $W_S$ is a quantity that characterizes strength of the NSID electron-nucleon interactions in the molecules. These two quantities cannot be measured, and are instead obtained from sophisticated molecular structure calculations in Refs. \cite{Meyer:2008gc, Dzuba:2011, Skripnikov:2013, Skripnikov:2015, Skripnikov:2016} for ThO and \cite {Petrov:2007, Fleig:2013, Skripnikov:2017, Fleig:2017} for HfF$^+$. The current values of the frequency are reported in \cite{Baron:2017,  Cairncross:2017fip}: 
\begin{align}
\omega_{\rm ThO}&=2.6\pm 4.8_{\rm stat}\pm3.2_{\rm syst}~{\rm mrad}/{\rm s},\hspace{0.8cm}\\
\omega_{\rm HfF}&=0.6\pm 5.4_{\rm stat}\pm 1.2_{\rm sys}~{\rm mrad}/{\rm s}.
\end{align} 

The electron EDM and the NSID interactions arise from new physics beyond the Standard Model (BSM).\footnote{The electron EDM is also generated by SM CPV interactions, but the corresponding magnitude is $\sim~10^{-38}~e~$cm \cite{Ng:1995cs} which is much smaller than the experimentally accessible value.} The NSID electron-nucleon interaction itself arises from semileptonic four-fermion interactions. Those most relevant to the paramagnetic systems are given by \cite{Engel:2013lsa}
\begin{align}
{\cal L}^{4F}=\frac{1}{\Lambda^2}\left[C_{ledq}O_{ledq}+C^{(1)}_{lequ}O^{(1)}_{lequ}\right],
\label{eq:fourfermion}
\end{align}
where 
\begin{align}
O_{ledq}&=\bar{L}^je_R\bar{d_R}Q^j, \label{4f_ledq} \\
O^{(1)}_{lequ}&=\bar{L}^je_R\epsilon_{jk}\bar{Q^k}u_R \label{4f_lequ},
\end{align}
with the $SU(2)$ indices $i,j$ and $k$, and
\begin{align}
L=
\begin{pmatrix}
\nu_L\\
e_L
\end{pmatrix},\hspace{0.5cm}
Q=
\begin{pmatrix}
u_L\\
d_L
\end{pmatrix}.
\end{align}
The coefficients in Eq. (\ref{NSID}) are expressed in terms of the Wilson coefficients appearing in Eq.~(\ref{eq:fourfermion}) as
\begin{align}
C^{(0)}_S&=-g^{(0)}_S\left(\frac{v}{\Lambda} \right)^2 {\rm Im}\left(C_{eq}^{-}\right),\\
C^{(1)}_S&=g^{(1)}_S\left(\frac{v}{\Lambda} \right)^2 {\rm Im}\left(C_{eq}^{+}\right),
\end{align}
with $C_{eq}^{\pm}\equiv C_{ledq}\pm C^{(1)}_{lequ}$. The quantities $g^{(0)}_S$ and $g^{(1)}_S$ represent isoscalar and isovector form factors, which are defined by
\begin{align}
\frac{1}{2}\left\langle N| \bar{u}u+\bar{d}d |N \right\rangle &=g^{(0)}_S\bar{\psi}_N\psi_N,\\
\frac{1}{2}\left\langle N| \bar{u}u-\bar{d}d |N \right\rangle &=g^{(1)}_S\bar{\psi}_N\tau^3\psi_N.
\end{align}
The values of these form factors are obtained in \cite{Engel:2013lsa}:
\begin{align}
g_S^{(0)}=6.3\pm0.8,\hspace{1cm}g_S^{(1)}=0.45\pm0.15.
\end{align}

\begin{table}[t]
\begin{tabular}{c | c | c | c }
\hline
System & $E_{\rm eff}$~[GV~cm$^{-1}$] & $W_S$~[kHZ]  & $\alpha_j~[e~$cm] \\
\hline \hline
$^{232}$Th$^{16}$O~~$(Z=90)$ & $78$  \cite{Baron:2017} & $-282$~\cite{Baron:2017} & $1.5\times 10^{-20}$ \\
$^{180}$Hf$^{19}$F$^+$~$(Z=72)$ & $-23$ \cite{Cairncross:2017fip} & $50.3$ \cite{Skripnikov:2017} & $9.0\times 10^{-21}$ \\
\hline
\end{tabular}
\caption{The values of $E_{\rm eff}$, $W_S$ and $\alpha_j$.}
\label{table:Wd_Ws}
\end{table}
\begin{figure}[t]
\begin{center}
\includegraphics[width=6cm]{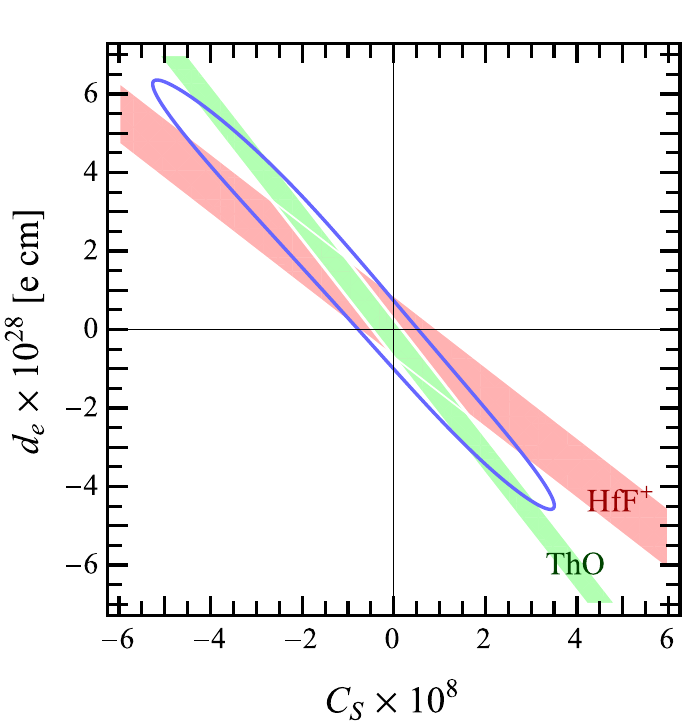} 
\end{center}
\caption{The allowed regions in $d_e$ and $C_S$ plane by the experimental results in ThO and HfF$^+$. The green and pink bands correspond to $d_{\rm ThO}$ and $d_{\rm HfF^+}$, respectively. The blue line is the contour with $90\%$ C.L. }
\label{GA}
\end{figure}

Following Refs. \cite{Chupp:2017rkp, Chupp:2014gka}, we define  an effective EDM $d_j$ for a given paramagnetic system as 
\begin{align}
d_j \equiv d_e+\alpha_{j}C_S,
\end{align}
where $j$ is either ThO or HfF$^+$. The quantity $\alpha_j$ is proportional to a ratio of $W_S/E_{\rm eff}$,\footnote{$\alpha_j$ corresponds to $\alpha_{C_S}/\alpha_{d_e}$ in Refs. \cite{Chupp:2017rkp, Chupp:2014gka}.} and
the values of $E_{\rm eff},~W_S$ and $\alpha_j$ are listed in Table~\ref{table:Wd_Ws}.
The experimental results can be translated into  the $d_j$ as 
\begin{align}
d_{\rm ThO}&=\left(-2.2\pm4.8\right) \times 10^{-29}~e~{\rm cm}, \label{d_ThO}\\
d_{\rm HfF^+}&=\left(0.9\pm7.9\right) \times 10^{-29}~e~{\rm cm}. \label{d_HfF}
\end{align}
In Fig. \ref{GA}, we recast a global analysis presented in Ref. \cite{Chupp:2017rkp}. For an analysis combining atomic EDMs, including the diamagnetic $^{199}$Hg system, see \cite{Fleig:2018bsf}.  The green band represents $d_{\rm ThO}$ with $1\sigma$ error, while the pink band is $d_{\rm HfF^+}$. The value of $W_S$ for HfF$^+$ is $2\pi$ times smaller than that in \cite{Chupp:2017rkp},\footnote{We thank Timothy Chupp for useful discussions on this point. } yielding the more tilted pink band.  The blue line corresponds to $90\%$ confidence-level contour, where we assume that theoretical uncertainties originating from $\alpha_j$ amount to $10\%$ as in \cite{Chupp:2017rkp}. The fit implies that
\begin{align}
-5\times 10^{-8} \lesssim~ &C_S~\lesssim3.5\times 10^{-8},\\
-4.5\times 10^{-28}~e~{\rm cm}\lesssim~ &d_e~ \lesssim 6.5\times 10^{-28}~e~{\rm cm}.
\end{align}
Inclusion of the $^{199}$Hg results provides  orthogonal constraints on $d_e$ and $C_S$, leading to somewhat tighter bounds \cite{Fleig:2018bsf}.  Given the anticipated future improvements in the sensitivities of the polar molecule experiments and their relatively simple interpretation, we concentrate here on ThO and HfF$^+$.

\section{Leptoquark Model}
\label{sec:model}
\begin{figure}[t]
\begin{center}
\includegraphics[width=9cm]{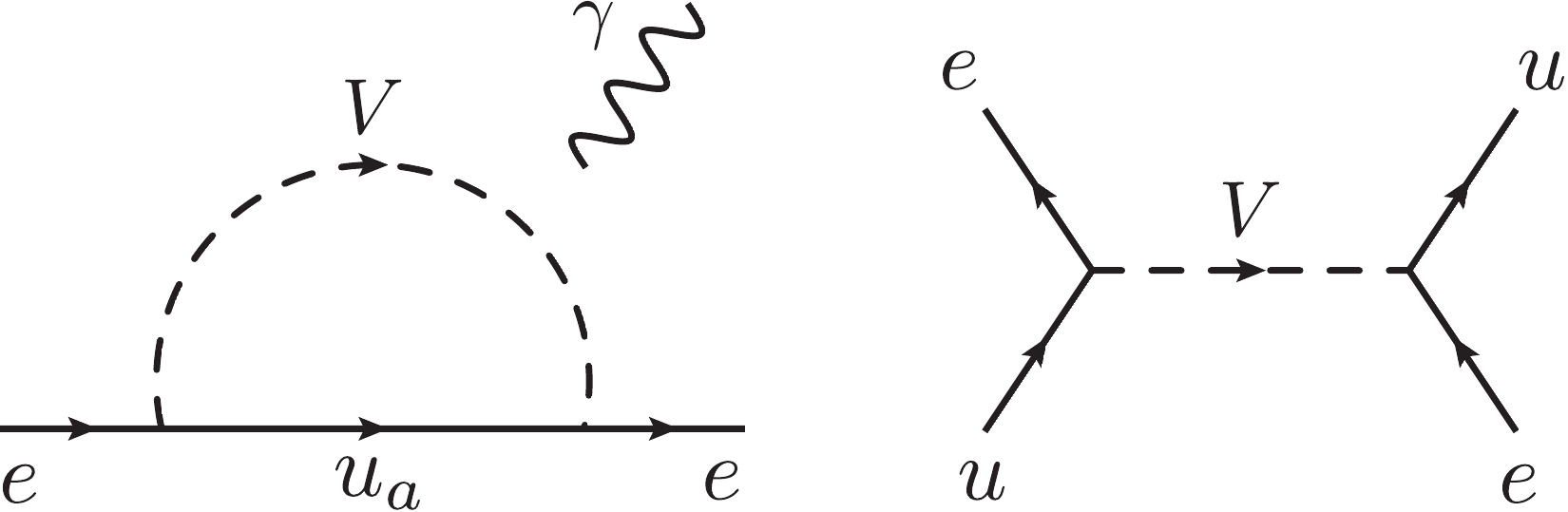} 
\end{center}
\caption{1-loop diagram of the electron EDM and the 4-fermi semileptonic operator.}
\label{fig:edm}
\end{figure}
A LQ is a field that can simultaneously couple to a quark and lepton at tree level, and possible SM gauge-invariant models are discussed in \cite{Buchmuller:1986zs, Davies:1990sc, Dorsner:2016wpm}. Here, we focus on scalar LQ's that have both left- and right-handed chiral couplings as needed to generate both the EDM and $eq$ operators. The only scalar LQ satisfying this requirement is the $X=({\bf 3},{\bf 2},7/6)$, where the numbers represent the SU(3)$_C\times$SU(2)$_L\times$U(1)$_Y$ representations. The couplings to the SM fermions are given by \cite{Arnold:2013cva}
\begin{align}
{\cal L}\ni -\lambda^{ab}_u\bar{u}^a_RX^T\epsilon L^b-\lambda^{ab}_e\bar{e}^a_RX^{\dagger}Q^b+{\rm h.c.}, \label{LQ_SM}
\end{align}
with
\begin{align}
X=
\begin{pmatrix}
V\\
Y
\end{pmatrix},
\end{align}
where $a$ and $b$ are flavor indices and $\epsilon$ is the antisymmetric tensor with $\epsilon_{12}=1$. Note that the presence of two distinct interactions in Eq.~(\ref{LQ_SM}) also allows for the presence of a relative phase between the two terms, an additional requirement for generation of the CPV operators of interest here. 

It should be noted that, if the LQ has a nonzero fermion number defined by $F=3B+L$, it can possess diquark couplings which results in proton decay at tree level. As discussed in \cite{Arnold:2012sd}, however, the scalar LQ $X$ does not induce this decay at tree level since $F=0$. In terms of the proton decay, another scalar LQ with quantum numbers of $({\bf 3},{\bf 2},1/6)$ does not have the diquark couplings, either. However, this LQ couples only to the lepton doublet and right-handed down-type quarks. Therefore, it cannot induce the chirality-flip  required for the EDM and CPV $eq$ interactions.

The left- and right-handed leptons in Eq.~(\ref{LQ_SM}) couple to the right- and left-handed up-type quarks, respectively. The interactions can induce the electron EDM at 1-loop order through the chirality flip of the up-type quarks as seen in left diagram of Fig. \ref{fig:edm}. The 1-loop diagram yields
\begin{align}
d_e=-\frac{em_{u_a}N_C}{32\pi^2m^2_V}{\rm Im}\left(\lambda^{a1}_u\lambda^{1a}_e \right)\left[Q_uI_2(X_{a})+Q_{LQ}J_2(X_a) \right]  \label{electron_EDM}
\end{align}
where the LQ mass $m_V$, $N_C=3$, $Q_u=2/3$, $Q_{\rm LQ}=5/3$, $X_a=m^2_{u_a}/m^2_V$. The loop functions $I_2(X_a)$ and $J_2(X_a)$ are given by
\begin{align}
I_2(X_a)&=\frac{1}{(1-X_a)^3}\left(-3+4X_a-X^2_a-2\log X_a \right),\\
J_2(X_a)&=\frac{1}{(1-X_a)^3}\left(1-X^2_a+2X_a\log X_a\right).
\end{align}
These results are in agreement with those in \cite{Bernreuther:1990jx,Baek:2015mea}.\footnote{The $I_2(X_a)$ function differs from that in \cite{Arnold:2012sd}.}
As seen in Eq. (\ref{electron_EDM}), the scale of $d_e$ is governed by the mass (or Yukawa coupling) of the intermediate quark. In principle, allowing a flavor non-diagonal coupling of the LQ to the electron and top quark would yield a contribution to $d_e$ nearly $10^{5}$ larger than obtained using the flavor diagonal coupling. However, as discussed below and in Ref.~\cite{Arnold:2012sd}, this possibility is severely constrained by the corresponding CP-conserving contribution to the electron mass and naturalness considerations. 

Similarly, the interactions in Eq. (\ref{LQ_SM}) lead to a non-vanishing up-quark EDM and chromo EDM that, in turn, contribute to  neutron and proton EDMs. The up-quark EDM is obtained by just replacing the up-quark parts in Eq. (\ref{electron_EDM}) with those of the electron and removing $N_C$. In addition, the chromo EDM contains only the $J_2$ loop function. The logarithmic part in the $I_2$ function governs the magnitude of the up-quark EDM in the neutron and proton EDMs. However, their sizes are roughly an order of magnitude smaller than $d_e$ as one would expect from the scaling with $N_C$ and virtual fermion Yukawa couplings:  $d_e/d_u \sim m_u N_C/m_e\sim 10$.

The LQ interactions also produce the 4-fermion semileptonic operators in Eq. (\ref{4f_lequ}) at tree level, as shown  in the right panel of Fig. \ref{fig:edm}. After a Fierz transformation, this amplitude yields 
\begin{align}
\frac{1}{\Lambda^2}{\rm Im}\left[C^{(1)}_{lequ}\right]=\frac{1}{2m^2_V}{\rm Im}\left(\lambda^{11}_u\lambda^{11}_e\right).
\end{align}
Note that the operator in Eq. (\ref{4f_ledq}) is not present since the LQ does not couple to the down-type quarks. And, it follows that $C_{eq}^{\pm}=\pm C^{(1)}_{lequ}$. In what follows, we parameterize Im$\left(\lambda^{a1}_u\lambda^{1a}_{e} \right)$ as $\left|\lambda^{a1}_u\lambda^{1a}_{e} \right|\sin\theta_{ue}$.

In principle, vector LQ interactions can also generate the electron EDM and the semileptonic four-fermion interactions. However, the vector LQ model requires an adequate UV completion to account for the origin of the vector LQ mass and to ensure renormalizability. Therefore, we focus exclusively on the scalar LQ case.

\section{Results}
\label{sec:results}
\begin{figure}[t]
\begin{center}
\includegraphics[width=6cm]{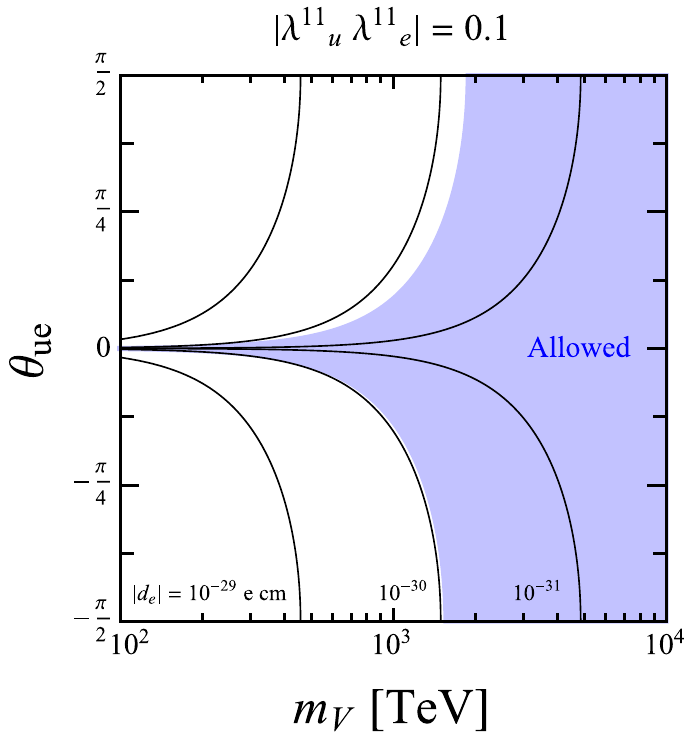}
\includegraphics[width=6cm]{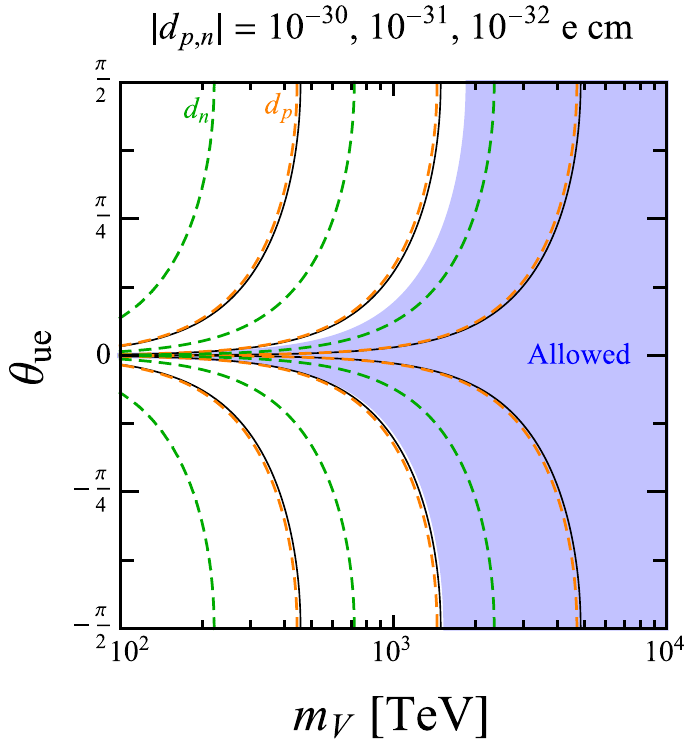}
\end{center}
\caption{(Upper) The allowed region with $90\%$ C.L in $(m_V,~\theta_{ue})$ plane with the fixed value of $\left|\lambda^{11}_u\lambda^{11}_e \right|=0.1$. The black lines are $|d_e|=1.0\times 10^{-29},~10^{-30}$ and $10^{-31}~e~$cm from left to right. (Lower) The proton (orange dashed line) and neutron (green dashed line) EDMs with $|d_{p,n}|=1.0\times 10^{-30},~10^{-31}$ and $10^{-32}~e~$cm from left to right.}
\label{fig:MFV01}
\end{figure}
We first focus on a situation where the LQ only interacts with the first generation leptons and quarks. In this case, the up quark runs in the loop in Fig. \ref{fig:edm}, and both the electron EDM and the semileptonic four-fermion operators are proportional to Im$\left(\lambda^{11}_u\lambda^{11}_e \right)$. In order to assess dependences on the CP phase $\theta_{ue}$ and the LQ mass $m_V$, we take $\left|\lambda^{11}_u\lambda^{11}_e\right|=0.1$ as a representative value. Figure~\ref{fig:MFV01} presents the allowed region at $90\%$ C.L. as implied by the ThO and HfF$^+$ results (the blue region). We also show contours  of constant $d_e$ (black lines) as a function of $m_V$ and $\theta_{ue}$. From left to right these contours represent $|d_e|=1.0\times 10^{-29},~10^{-30},~10^{-31}~e~$cm.
The boundary of the blue region in the negative $\theta_{ue}$ region corresponds to $|d_e|\simeq 1.0\times 10^{-30}~e~$cm. In short, the constraints on $m_V$ and Im$\left(\lambda^{11}_u\lambda^{11}_{e} \right)$ are dominated by those on the NSID interaction, and imply an upper bound on $d_e$ that is two orders of magnitude smaller than the limit obtained with the sole-source assumption. 
As a corollary for the LQ scenario therefore $d_j$ can be described by
\begin{align}
d_j &\sim \alpha_jC_S\nonumber\\
&\sim O(1)\times\left(\frac{\left|\lambda^{11}_u\lambda^{11}_e \right|\sin\theta_{ue} }{0.1} \right) \left(\frac{1000~{\rm TeV}}{m_V} \right)^2 \times 10^{-28}. \nonumber
\end{align}

The lower plot in Fig. \ref{fig:MFV01} reflects the proton and neutron EDMs with orange and green dashed lines. Here, we employ formulae for them in \cite{Hisano:2015rna}.  Their magnitudes are $|d_{p,n}|=1.0\times 10^{-30},~10^{-31}$ and $10^{-32}~e~$cm from left to right. As discussed in the previous section, they become an order of magnitude smaller than the electron EDM. In addition, since $d_p/d_n \sim 0.8d_u/0.2d_u\sim 4$, $d_n$ is somewhat smaller than $d_p$.

\begin{figure}[t]
\begin{center}
\includegraphics[width=6cm]{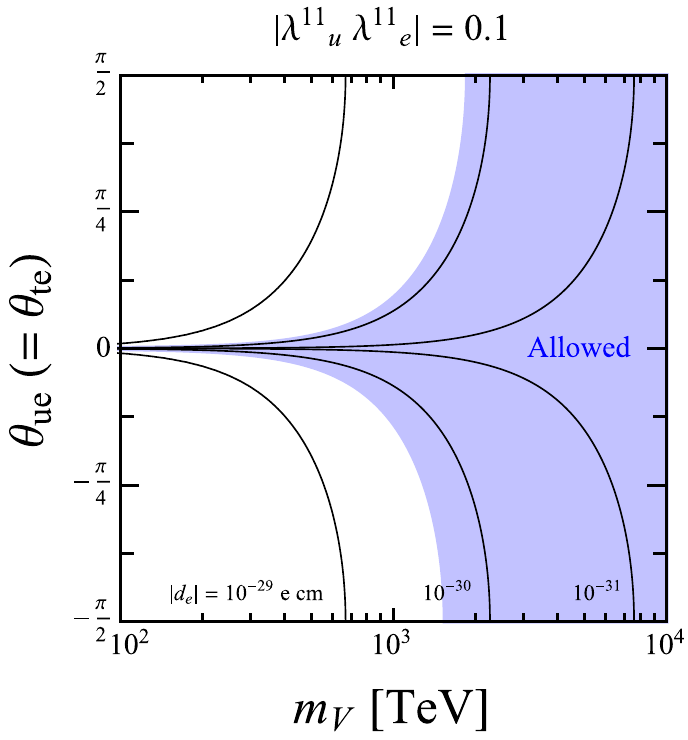}
\end{center}
\caption{The allowed region in $(m_V,~\theta_{ue})$ plane including the top-quark contribution with nonzero $\lambda^{31}_u\lambda^{13}_e$. It is taken that $\left|\lambda^{11}_u\lambda^{11}_e \right|=0.1$ and $\theta_{ue}=\theta_{te}$. The black contours represent the electron EDM.}
\label{fig:FV01}
\end{figure}
Next, we introduce the LQ interaction to the top quark. In this case, the top quark can also contribute to the 1-loop EDM via the left panel in Fig. \ref{fig:edm}. This contribution is proportional to $\left|\lambda^{31}_u\lambda^{13}_e \right|$. There is no corresponding tree-level contribution to the NSID interaction, which involves only electrons and first generation quarks. As discussed in \cite{Arnold:2013cva}, the flavor non-diagonal interaction can also generate a contribution to the electron mass that is enhanced by a factor of $m_t$: 
\begin{align}
\Delta m_e \simeq \left|\lambda^{31}_u\lambda^{13}_e\right|\frac{3m_t}{16\pi^2}\log\left(\frac{\Lambda^2}{m^2_V}\right),
\label{eq:me}
\end{align}
where $\Lambda$ is a cutoff scale and where we use the leading log approximation.  The $\Lambda$-independence of $m_e$ requires the presence of a corresponding counter term. 

The presence of the $m_t$-enhancement in Eq.~(\ref{eq:me}) requires a fine-tuned cancellation between the finite parts of the one-loop and counter term contributions, unless $\left|\lambda^{31}_u\lambda^{13}_e\right|$ is sufficiently small. To illustrate, we 
take the GUT scale as the cutoff scale, $\Lambda=1.0\times 10^{16}~$GeV, and $m_V=10^3~$TeV which yields $\Delta m_e \simeq \left|\lambda^{31}_u\lambda^{13}_e\right| \times 151$ GeV.\footnote{In the case of the up quark, its contribution to $\Delta m_e$ is roughly $O(10^{-4})$ GeV with $\left|\lambda^{11}_u\lambda^{11}_e\right|=0.1$.}  In order to avoid this large contribution, we require that the contribution is the same order of magnitude as the electron mass at most. This \lq\lq naturalness" condition implies that $\left|\lambda^{31}_u\lambda^{13}_e\right| = m_e/\left[{3m_t}/{16\pi^2}\log\left({\Lambda^2}/{m^2_V}\right)\right]$, which leads to $\left|\lambda^{31}_u\lambda^{13}_e\right| \sim O(10^{-6})$ for the range of $m_V$ considered here. In what follows, we employ this condition to determine the magnitude of $\lambda^{31}_u\lambda^{13}_e$.

Figure \ref{fig:FV01} shows the allowed region and the size of the electron EDM as in Fig. \ref{fig:MFV01} including both the top-quark and up-quark contributions. Here, we take that $\left|\lambda^{11}_u\lambda^{11}_e \right|=0.1$ and $\theta_{ue}=\theta_{te}$. Although the top-quark mass is expected to enhance the electron EDM, the naturalness requirement on the coupling compensates for the enhancement. In the current setup, it turns out that $m_u\left|\lambda^{11}_u\lambda^{11}_e \right|/m_t\left|\lambda^{31}_u\lambda^{13}_e \right|\sim 1$, and the electron EDM roughly becomes just two times larger than the size in the previous case. Therefore, the dominant contribution to the polar molecule systems remains the NSID interaction. The additional contribution to $d_e$ somewhat shifts the black contours to right, which results in $|d_e| \lesssim 2.0\times 10^{-30}~e~$cm.  It is, of course, possible that the aforementioned naturalness considerations are not realized, in which case the flavor non-diagonal contribution to $d_e$ could be considerably larger. 
Moreover, if only the flavor non-diagonal interactions are allowed, the NSID interaction is not present. Therefore, the well-known sole-source limit applies to this case.

\section{Conclusion}
\label{sec:conclusion}
In recent years, bounds of the electron EDM have been extraordinarily improved by utilizing polar molecules  ThO and HfF$^+$. These molecular systems admit significantly larger internal electric fields than one can produce in the laboratory, a feature that significantly enhances the sensitivity to the electron EDM. The corresponding $d_e$ limits typically assume that only the electron EDM contributes to the P- and T-violating effect -- an  assumption that is valid as long as the NSID electron-nucleon interaction is suppressed.

Leptoquark models provide one exception to this scenario, as LQ interactions may simultaneously induce both the EDM and NSID interactions. Moreover,  the latter arise at tree-level without a Yukawa insertion, whereas the EDM first appears at one-loop order and requires a Yukawa coupling as implied by gauge invariance. From the general considerations outlined in Ref.~\cite{Chupp:2014gka}, we  expect the LQ NSID contribution to dominate the ThO and HfF$^+$ responses. From our explicit study, we find that for fixed values of the LQ couplings, the corresponding lower bounds on the LQ mass implied by the NSID interaction are at least an order of magnitude stronger than those associated with the electron EDM. Consequently, the magnitude of $d_e$ must be nearly two orders of magnitude smaller than implied by a sole-source analysis of the experimental results. 

It is instructive to translate these conclusions into expectations for the neutron and proton EDMs. In the LQ scenario, the nucleon EDMs are suppressed relative to $d_e$ by the ratio of the electron and light-quark masses as well as by $1/N_C$. From a numerical study, we find that the present paramagnetic molecular results imply magnitudes for $d_n$ and $d_p$ roughly comparable to the expected Standard Model, CKM contributions \cite{Gavela:1981sk, Khriplovich:1981ca, McKellar:1987tf, Seng:2014lea}.
Thus, the observation of a non-vanishing neutron or proton EDM in future experiments would point to a different source of CPV than associated with LQ interactions. Conversely, a non-vanishing signal in the next generation paramagnetic EDM searches would be consistent with a LQ-induced NSID interaction.

\begin{acknowledgments}
We are grateful to Timothy Chupp for useful discussions and Vincenzo Cirigliano for detailed analytic results on the leptoquark interactions. KF thanks Hao-Lin Li, Hiren Patel and Jiang-Hao Yu for several helpful conversations.
This work was supported in part under U.S. Department of Energy contract DE-SC0011095.
\end{acknowledgments}


\end{document}